%% file: main.tex
\documentclass[aps,prl,10pt,tightenlines,notitlepage,showpacs,nofootinbib,superscriptaddress,twocolumn,raggedbottom,preprintnumbers,longbibliography]{revtex4-1}

\usepackage{amsfonts}
\usepackage{amsmath}
\usepackage{amssymb}
\usepackage{array}
\usepackage{bbold}
\usepackage{bm}
\usepackage{booktabs}
\usepackage{dsfont}
\usepackage{enumitem}
\usepackage{float}
\usepackage{graphicx}
\usepackage[utf8]{inputenc}
\usepackage{lastpage}
\usepackage{nicefrac}
\usepackage[normalem]{ulem}
\usepackage[dvipsnames]{xcolor}
\usepackage{hyperref}
\hypersetup{
    colorlinks=true,
    urlcolor=blue,
    linkcolor = blue,
    urlcolor  = blue,
    citecolor = blue,
    anchorcolor = blue,
    pdftitle={a1 pheno},
    pdfauthor={friends},
    pdfsubject={a1}}

\renewcommand{\Re}{{\rm Re}}

\input{chris_definitions.tex}

\begin{document}


\title{
Three-body dynamics of the \texorpdfstring{$a_1(1260)$}{a1(1260)} resonance from lattice QCD
}
\author{Maxim~Mai}
\email{maximmai@gwu.edu}
\affiliation{The George Washington University, Washington, DC 20052, USA}
\author{Andrei~Alexandru}
\email{aalexan@gwu.edu}
\affiliation{The George Washington University, Washington, DC 20052, USA}
\affiliation{Department of Physics, University of Maryland, College Park, MD 20742, USA}
\author{Ruair\'{i}~Brett}
\email{rbrett@gwu.edu}
\affiliation{The George Washington University, Washington, DC 20052, USA}
\author{Chris~Culver}
\email{C.Culver@liverpool.ac.uk}
\affiliation{Department of Mathematical Sciences, University of Liverpool, Liverpool L69 7ZL, United Kingdom}
\author{Michael~D\"oring}
\email{doring@gwu.edu}
\affiliation{The George Washington University, Washington, DC 20052, USA}
%
\author{Frank~X.~Lee}
\email{fxlee@gwu.edu}
\affiliation{The George Washington University, Washington, DC 20052, USA}
\author{Daniel Sadasivan}
\email{daniel.sadasivan@avemaria.edu}
\affiliation{Ave Maria University, Ave Maria, FL 34142, USA}

\collaboration{GWQCD Collaboration}

\begin{abstract}
Resonant hadronic systems often exhibit a complicated decay pattern in which three-body dynamics play a relevant or even dominant role.
In this work we focus on the $a_1(1260)$ resonance.
For the first time, the pole position and branching ratios of a three-body resonance are calculated from lattice QCD using one-, two-, and three-meson interpolators and a three-body finite-volume formalism extended to spin and coupled channels.
This marks a new milestone for ab-initio studies of ordinary resonances along with hybrid and exotic hadrons involving three-body dynamics.
\end{abstract}
\maketitle


\noindent
\textit{Introduction} ---
Many unresolved questions in the excited spectrum of strongly interacting particles are related to the hadronic three-body problem~\cite{ParticleDataGroup:2020ssz}. Some examples of interest include: axial mesons like the $I^G(J^{PC})=1^-(1^{++})~a_1(1260)$ or exotic mesons, such as the $J^{PC}=1^{-+}$ $\pi_1(1600)$ claimed by the COMPASS collaboration~\cite{Alekseev:2009aa} by analyzing three-pion final states; this and other exotic mesons searched for in the GlueX experiment~\cite{Dobbs:2019sgr}; the Roper resonance $N(1440)1/2^+$ with its unusually large branching ratio to the $\pi\pi N$ channel and a very non-standard line shape~\cite{Arndt:2006bf,Ceci:2011ae,Loring:2001kx,Lang:2016hnn}; heavy mesons like the $X(3872)$ with large branching ratio to
$D\bar D\pi$ states~\cite{Belle:2003nnu, Belle:2006olv}. Furthermore, multi-neutron forces are crucial for the equation of state of a neutron star~\cite{Baym:2017whm}. Recent advances in lattice QCD (LQCD) on few-nucleon systems~\cite{Beane:2012vq,Savage:2016egr} complement dedicated experimental programs, e.g., at the FRIB facility~\cite{Geesaman:2015fha}.

Lattice QCD provides information about the structure and interactions
of hadrons as they emerge from quark-gluon dynamics. For scattering
this information is extracted indirectly by accessing the energy
of the multi-hadron states in finite volume. The connection to
infinite-volume scattering amplitudes is provided by {\em quantization
conditions}. In the two-hadron sector this technique is already a precision tool for extracting
phase-shifts and resonance information~\cite{Briceno:2017max, Detmold:2019ghl, Lang:2007mq, Doering:2014fpa, Briceno:2014tqa}. Moving to the three-hadron
sector new challenges emerge, both in terms of determining precisely
the energy of three-particle states from QCD and in
developing the necessary quantization conditions.

Three-hadron LQCD calculations have been performed mostly for pion and kaon systems at maximal isospin~\cite{Detmold:2008fn, Detmold:2008yn, Horz:2019rrn, Culver:2019vvu, Fischer:2020jzp, Hansen:2020otl, Alexandru:2020xqf, Blanton:2021llb}. Through the use of a large basis of one-, two-, and three-meson interpolators, these calculations provide reliable access to the energies of three-particle states and, using recently developed quantization conditions, infinite-volume amplitudes can be accessed~\cite{Polejaeva:2012ut, Briceno:2012rv, Roca:2012rx, Bour:2012hn, Meissner:2014dea, Jansen:2015lha, Hansen:2014eka, Hansen:2015zga, Hansen:2015zta, Hansen:2016fzj, Guo:2016fgl, Konig:2017krd, Hammer:2017uqm, Hammer:2017kms, Briceno:2017tce, Sharpe:2017jej, Guo:2017crd, Guo:2017ism, Meng:2017jgx, Guo:2018ibd, Guo:2018xbv, Klos:2018sen, Briceno:2018mlh, Briceno:2018aml, Mai:2017bge, Mai:2018djl, Doring:2018xxx, Jackura:2019bmu, Mai:2019fba, Guo:2019hih, Blanton:2019igq, Briceno:2019muc, Romero-Lopez:2019qrt, Pang:2019dfe, Guo:2019ogp, Zhu:2019dho, Pang:2020pkl, Hansen:2020zhy, Guo:2020spn, Guo:2020wbl, Guo:2020ikh, Guo:2020kph, Blanton:2020gha, Blanton:2020gmf, Muller:2020vtt, Brett:2021wyd, Muller:2020wjo, Hansen:2021ofl, Blanton:2021mih}. Among these approaches we highlight \emph{Relativistic Field Theory} (RFT)~\cite{Hansen:2014eka, Hansen:2015zga}, \emph{Non-Relativistic Effective Field Theory} (NREFT)~\cite{Hammer:2017uqm, Hammer:2017kms}, and \emph{Finite Volume Unitarity} (FVU)~\cite{Mai:2017bge, Mai:2018djl}. For reviews see Refs.~\cite{Mai:2021lwb, Hansen:2019nir, Rusetsky:2019gyk}.

So far, no resonant three-body system has been studied using any finite-volume methodology. In this letter we take on this challenge, calculating the excited-state spectrum of the $a_1(1260)$ in LQCD and subsequently mapping it to the infinite volume.  This enables, for the first time, the determination of resonance pole position and branching ratios for a three-body resonance from first principles.

The $a_1(1260)$  decays  exclusively to three pions~\cite{ParticleDataGroup:2020ssz, Alekseev:2009aa} and can be measured cleanly in $\tau$-decays~\cite{CLEO:1999rzk,Schael:2005am} allowing for its three-body decay channels to be determined. The resonance is wide~\cite{ParticleDataGroup:2020ssz} indicating strong and non-trivial three-body effects which make it a prime candidate to study three-body dynamics. This is reflected in an increased interest in the dynamics and the structure of the $a_1(1260)$~\cite{Janssen:1993nj, Lutz:2003fm, Geng:2006yb, Wagner:2007wy, Wagner:2008gz, Lutz:2008km, Kamano:2011ih, Nagahiro:2011jn, Zhou:2014ila, Zhang:2018tko, Dai:2018zki, Mikhasenko:2018bzm, Sadasivan:2020syi, Dai:2020vfc, Dias:2021upl} including pioneering calculations~\cite{Lang:2014tia, Roca:2005nm}. Of these approaches, Refs.~\cite{Kamano:2011ih, Sadasivan:2020syi} use frameworks that manifestly incorporate three-body unitarity which is the linchpin of the FVU formalism~\cite{Mai:2017bge} and a prerequisite for the mapping between finite and infinite volume.

We generalize the FVU formalism to include two-particle subsystems with spin, to map the LQCD spectrum to the resonance pole of the $a_1(1260)$. Furthermore, the dominant decay of the $a_1(1260)$ into $\pi\rho$ occurs in two channels (S/D-wave) which requires an upgrade of the formalism to coupled channels. Finally, the challenge of analytic continuation of three-body amplitudes to complex pole positions is also resolved in this study and we deliver the first three-body unitary pole determination of the $a_1$ from experiment.

By calculating the excited LQCD spectrum, mapping it to the infinite-volume coupled-channel amplitude, and finally determining the $a_1(1260)$ pole and branching ratios we demonstrate that detailed calculations of three-body resonances from first principles QCD have become possible. This paves the way for the ab-initio understanding of a wide class of resonance phenomena, including hybrid and exotic hadrons, that lie at the heart of non-perturbative QCD.

\bigskip
\noindent
\textit{LQCD spectrum} ---
We extract the finite-volume spectrum in the $a_1(1260)$ sector using an ensemble with $N_f=2$ dynamical fermions, with masses tuned such that the pion mass is $224~\text{MeV}$.
The lattice spacing $a=0.1215~\text{fm}$ is determined using Wilson flow parameter $t_0$~\cite{Niyazi:2020erg}.  This ensemble has been used multiple times~\cite{Pelissier:2012pi,Guo:2016zos,Guo:2018zss,Culver:2019qtx,Culver:2019vvu,Alexandru:2020xqf} to successfully study two- and three-meson scattering, thus, we will only review the most important calculation details and new features relevant for the $a_1(1260)$. Computationally expensive quark propagators are estimated with LapH smearing~\cite{Peardon:2009gh}, calculated using an optimized inverter~\cite{Alexandru:2011ee}.  Having access to the so-called perambulators makes it straightforward to construct a large basis of operators for use in the variational method~\cite{Michael:1982gb,Luscher:1990ck,Blossier:2009kd}, which removes excited state contamination and allows extraction of the excited state spectrum.

Performing the calculation in a cubic volume reduces the rotational symmetry group $SO(3)$ to the group $O_h$. States on the lattice thus cannot be classified by their angular momentum quantum number. Instead, they are classified by the irreducible representations~(irreps) of $O_h$.  For the $a_1(1260)$ the irrep of interest is $T_{1g}$ which subduces onto the continuum quantum numbers $J^{P}=1^{+}$.  Aside from ensuring that our operators have the correct angular momentum content we must also construct them to have total isospin $I=1$ to match the $a_1(1260)$.
The last major consideration for constructing our operator basis is to ensure sufficient overlap with the lowest-lying states of the spectrum. In that regard we utilize both a single-meson $\bar{q}q$ operator and multi-meson operators for each of the most prominent decay channels of the $a_1(1260)$, $\rho\pi, \sigma\pi$, and $\pi\pi\pi$.  Further details of the operator construction can be found in supplement~\ref{APP:lat}~\cite{supp}.

The most challenging aspect of the calculation is ensuring the operator basis is sufficient to extract the states below the inelastic scattering threshold.  For this ensemble and symmetry channel, there are only two such states, yet 11 operators were required to stabilize the fit of the first excited state. We find that the stability of the excited state relies heavily on the inclusion of a three-pion operator where two of the pions have back-to-back momenta $\frac{2\pi}{L}(1,1,0)$, despite the expected non-interacting energy of such a three-pion state lying far above the inelastic threshold. In addition, we also ensure stability under the variation of fit range and variational parameters. The obtained energy eigenvalues are depicted in Fig.~\ref{fig:fit-to-levels}, see supplement~\ref{APP:statistics}~\cite{supp} for numerical values.

\bigskip
\noindent
\textit{Quantization condition} ---
The $a_1(1260)$ couples to  three-pion states in the $I^G(J^{PC})=1^-(1^{++})$ channel that can be decomposed as $\pi\rho$ in S/D-wave, $\pi f_0(500)$ and $\pi (\pi\pi)_{I=2}$ in P-waves and other channels. Phenomenologically $(\pi\rho)_S$ is dominant~\cite{Kuhn:2004en} with the branching ratios into other channels quite uncertain~\cite{ParticleDataGroup:2020ssz}.

Since the isoscalar $\pi\pi$ interaction weakens at heavier pion mass~\cite{Mai:2019pqr,Guo:2018zss,Doring:2016bdr}, for now we restrict the discussion to the $\pi\rho$ channels. In that, and following the unitary three-body formalism~\cite{Mai:2017vot,Sadasivan:2020syi}, the $\pi(p_1)\pi(p_2)\pi(p_3)\to \pi(p_1')\pi(p_2')\pi(p_3')$ scattering amplitude can be re-written in terms of a two-pion spin-1 cluster, carrying a helicity index $\lambda^{(\prime)}\in\{-1,0,1\}$, and a third pion (spectator). For $s:=(p_1+p_2+p_3)^2=:P^2,~\sigma_l:=(P-l)^2~\text{and}~E_l:=\sqrt{\bm{l}^2+m_\pi^2}$ this yields
\begin{widetext}
\begin{align}
\nonumber
&\langle p'_1p'_2p'_3|T_3(s)|p_1p_2p_3\rangle=
\aleph \sum_{\substack{\lambda,\lambda'\\m,n}}
\hat v_{\lambda'}(\bm{p}'_{\bar{n}},\bm{p}'_{\bar{\bar{n}}})
\Big(
\tau(\sigma_{p_n^\prime}) T_{\lambda'\lambda}^{c}(s,\bm{p}_n',\bm{p}_m)
+
2E_{p_n}(2\pi)^3\delta^3(\bm{p}_n'-\bm{p}_m)
\Big)
\tau(\sigma_{p_m})
\hat v_{\lambda}(\bm{p}_{\bar{m}},\bm{p}_{\bar{\bar{m}}}) \ ,\\
&T^c_{\lambda'\lambda}(s,{\bm p}',\bm{p})=
B_{\lambda'\lambda}(s,{\bm p}',\bm{p})+
C_{\lambda'\lambda}(s,{\bm p}',\bm{p})+
\int\frac{d^3\bm l}{(2\pi)^3 2E_l}
\big(
B_{\lambda'\lambda''}(s,{\bm p}',{\bm  l })+C_{\lambda'\lambda''}(s,{\bm p}',{\bm  l })
\big)
\tau(\sigma_l) T^c_{\lambda''\lambda}(s,{\bm l},{\bm{p}})\,,
\label{eq:T3-integral-equation}
\end{align}
\end{widetext}
where $\aleph$ is an isospin combinatorial factor, and
in each occurrence $\bar x\in\{1,2,3\}\backslash \{x\}$ and $\bar{\bar x}\in\{1,2,3\}\backslash \{x,\bar x\}$. The coupling of the spin-1 system to the asymptotic states is facilitated via $\hat v(p,q)_\lambda=-i\varepsilon_\lambda^\mu(p+q)\,(p_\mu-q_\mu)$ for the usual helicity state vectors $\varepsilon$, provided for convenience in supplement~\ref{APP:QC-details}~\cite{supp}. The $\pi\rho$ interaction kernel projected to $I=1$ consists of: 1) the one-pion-exchange term
\begin{align}
B_{\lambda'\lambda}(s,\bm{p}',\bm{p})=\frac{\hat v^*_{\lambda'}(P-p-p',p)\hat v_{\lambda}(P-p-p',p')}{2E_{\bm{p}'+\bm{p}}(\sqrt{s}-E_{\bm{p}}-E_{\bm{p}'}-E_{\bm{p}'+\bm{p}})}\,,
\label{eq:Bterm}
\end{align}
which is a consequence of three-body unitarity~\cite{Mai:2017vot}; and 2) a short-range three-body force generically parametrized by a Laurent series in the $JLS$ basis (${\ell^{(\prime)} \in\{S,D\}}$),
\begin{align}
C_{\ell'\ell}(s,\bm{p}',\bm{p})&=\sum_{i=-1}^\infty c_{\ell'\ell}^{(i)}(\bm{p}',\bm{p})(s-m_{a_1}^2)^i\,,
\label{eq:C-term}
\end{align}
including first-order poles to account for resonances. The projection to helicity basis follows standard procedure~\cite{Chung:1971ri}, recapitulated in supplement~\ref{APP:QC-details}~\cite{supp}.

The spin-1 propagator ensures two-body unitarity in all sub-channels and is expressed in terms of an $n$-times subtracted self-energy $\Sigma_n$ and a $K$-matrix-like quantity $\tilde K_n$,
\begin{align}
\label{eq:tau-infinite}
&\tau_{\lambda'\lambda}^{-1}(\sigma_p)=
\delta_{\lambda'\lambda}\tilde{K}_n^{-1}(s,\bm{p})-\Sigma_{n,\lambda'\lambda}(s,\bm{p})
\,,\\\nonumber
&\tilde{K}_n^{-1}(s,\bm{p})=\sum_{i=0}^{n-1} a_i\sigma_p^i
\quad\text{and}\quad
\Sigma_{n,\lambda'\lambda}(s,\bm{p})=\\\nonumber
&\int\frac{d^3k}{(2\pi)^3}
\frac{\sigma_p^n}{(4E_{k}^2)^n}
\frac{
\hat v^*_{\lambda'}(P-p-k,k)
\hat v_{\lambda}(P-p-k,k)
}{ 2E_{k}(\sigma_p-4E_{k}^2+i\epsilon)}\,.
\end{align}
We found that $n=2$ is sufficient to render the self-energy term convergent without destroying analytic properties of the amplitude in Eq.~\eqref{eq:T3-integral-equation}.

Putting an interacting multi-hadron system into a cubic box of size $L$ restricts the momentum space  $\mathds{R}^3\to\mathcal{S}_L:=(2\pi/L)\mathds{Z}^3$. This means that the integral equation~\eqref{eq:T3-integral-equation} becomes an algebraic one via $\int d^3\bm{k}/(2\pi)^3\to 1/L^3\sum_{\bm{k}\in{\mathcal{S}_L}}$, the solutions of which are singular iff mesons are on-shell. Thus, the positions of singularities in $s<5m_\pi$ are equivalent to the energy eigenvalues up to $e^{-m_\pi L}$ terms, determined from
\begin{align}
\label{eq:3bQC}
0=
{\rm det}\,\Big[B(s)+C(s)
-E_L
\left(
\tilde K_2^{-1}(s)-\Sigma_2^L(s)
\right)
\Big]_{\substack{(\lambda'\lambda)\\(\bm{p}'\bm{p})}}\,,
\end{align}
which defines the generalized FVU quantization condition. Here $E_L:=2E_pL^3$, while the explicit expression for the finite-volume $\Sigma^L_2$ is provided in supplement~\ref{APP:QC-details}~\cite{supp}. The major novelty induced by the $\rho$ spin lies in the non-diagonal $\Sigma_{\lambda'\lambda}$ corresponding to in-flight mixing of $\rho$ helicities.

We note that the determinant is taken over helicity and spectator momentum spaces. Finding the energies associated with a particular row $\mu$ of irrep $\Lambda$ of the symmetry group $G$ can be done in the standard fashion by block diagonalizing the quantization condition and examining the determinant only for the relevant block/irrep~\cite{Morningstar:2017spu,Brett:2021wyd}. In practice this is accomplished by first converting from the helicity basis to canonical state vectors, $\ket{\bm{p} \lambda} \rightarrow \ket{\bm{p} m}$, then block diagonalizing, $\ket{\bm{p} m}\rightarrow \ket{\lambda\mu}$.


\begin{figure}[tb]
    \centering
    \includegraphics[width=\linewidth]{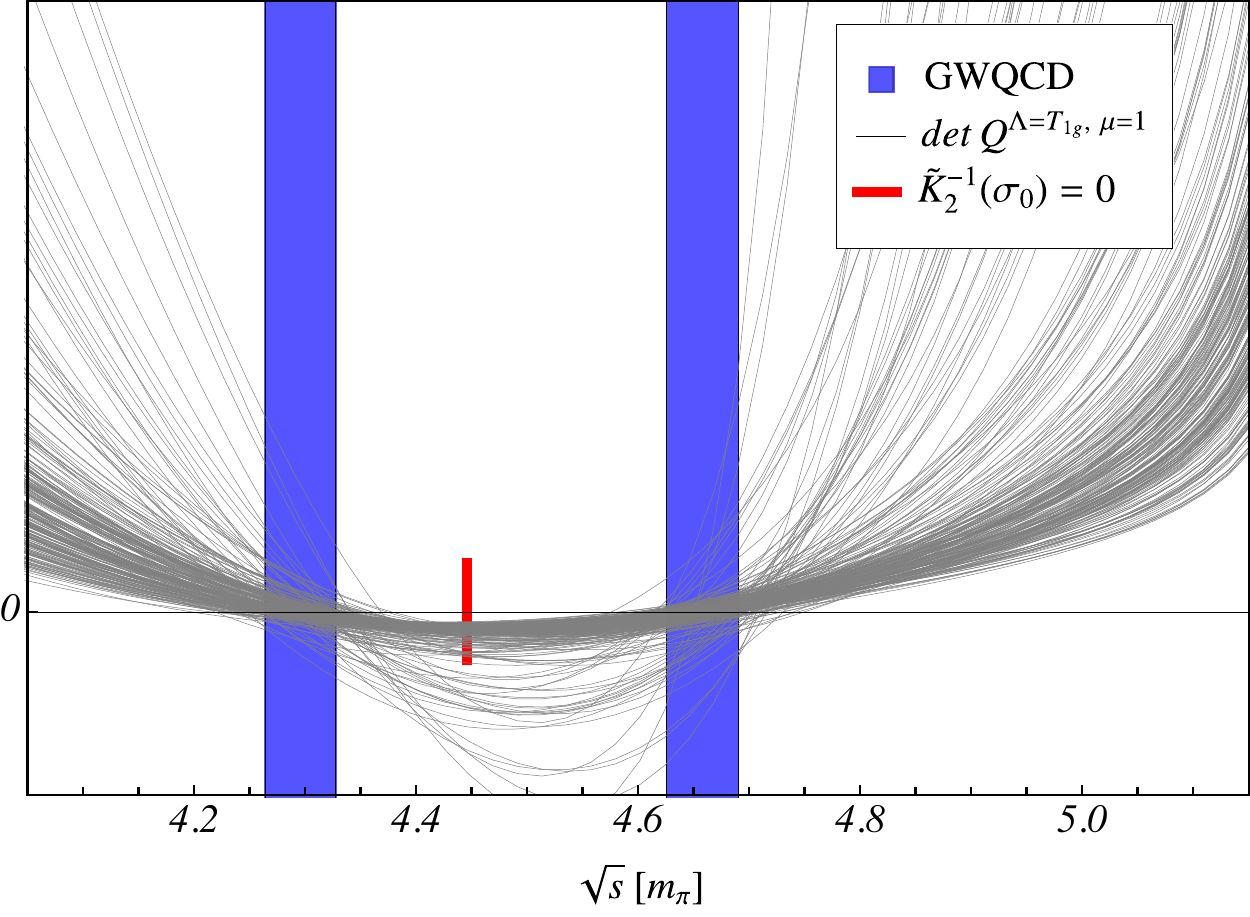}
    \caption{Right hand side of the quantization condition~\eqref{eq:3bQC} (gray) re-fitted to the correlated LQCD energy eigenvalues (blue bars indicating 1-$\sigma$ uncertainties). The red bar shows the position of the ground level in the case of vanishing interactions.}
    \label{fig:fit-to-levels}
\end{figure}

\bigskip
\noindent
\textit{Fits} ---
The quantization condition in Eq.~\eqref{eq:3bQC} contains the volume-independent, regular quantities $C$ and $\tilde K_2^{-1}$. We fix the parameters of the latter by using the two-pion finite-volume spectrum~\cite{Guo:2016zos,Guo:2018zss,Culver:2019qtx}, matching the isovector amplitude $T^{I=\ell=1}_{22}=\hat v\tau \hat v$ to the one determined in Ref.~\cite{Mai:2019pqr}. We obtain $a_0=-0.1577\,m_\pi^2,~a_1=0.0133$.

The three-body force in Eq.~\eqref{eq:C-term} is inherently cutoff-dependent with respect to the spectator momentum in Eq.~\eqref{eq:3bQC}. This cutoff needs to be held fixed when connecting finite and infinite-volume quantities. We take $|\bm{p}|\leq 2\pi/L|(1,1,0)|\approx 2.69~m_\pi$. Finally, exploring various possibilities we found that truncating the general expansion \eqref{eq:C-term} according to
\begin{align}
C_{\ell' \ell}(s,\bm{p}',\bm{p})=
g_{\ell'}\left(\frac{|\bm{p}'|}{m_\pi}\right)^{\ell'}
\frac{m_\pi^2}{s-m_{a_1}^2}
g_{\ell}\left(\frac{|\bm{p}|}{m_\pi}\right)^\ell
+c\,\delta_{\ell' 0}\delta_{\ell 0}\,,
\label{eq:C-pars}
\end{align}
yields a sufficient parametrization of the three-body spectrum. We emphasize that with only two three-body levels (as expected for the given $m_\pi L\approx 3.3$) the fit parameters will be strongly correlated.

To assess statistical uncertainty, we perform fits of $\{m_{a_1},g_S,g_D,c\}$ to re-sampled energy eigenvalues, each time picking a random starting value $5\,m_\pi<m_{a_1}<12\,m_\pi$. The result is depicted in Fig.~\ref{fig:fit-to-levels} using a subset of all considered samples (2000). The distribution of parameters and correlations, along with $\chi^2$-distributions are provided in supplement~\ref{APP:statistics}~\cite{supp}. We find the largest correlations in $(m_{a_1},g_D)$ and $(g_S,c)$,
meaning that the bare mass $m_{a_1}$ can be easily renormalized by the D-wave $a_1$ self energy which is proportional to $g_D^2$; indeed, the latter is almost real in the considered energy region and therefore strongly correlated with the real $m_{a_1}$ parameter. As a sanity check, when adiabatically tuning down the $\pi\pi$ interaction, the ground level indeed approaches the energy at which $\tilde K^{-1}_2=0$ (red bar in Fig.~\ref{fig:fit-to-levels}), as the $\rho$ becomes infinitely narrow and stable at the corresponding invariant mass.

\bigskip
\noindent
\textit{Analytic continuation and poles} ---
To extract the physical resonance parameters, i.e., the pole position and branching ratios of the $a_1(1260)$, we turn back to the infinite-volume scattering amplitude in Eq.~\eqref{eq:T3-integral-equation}. With all parameters fixed from the lattice, $T^c_{\ell'\ell}$ is calculated in the $JLS$ basis~\cite{Sadasivan:2020syi,Sadasivan:2020tmy}. The integration over spectator momenta is performed on a complex contour, avoiding singularities for both real and complex-valued $\sqrt{s}$. See supplement~\ref{APP:projections}~\cite{supp} for technical
details and the projection $T^c_{\lambda'\lambda} \rightarrow T^c_{\ell'\ell}$.

\begin{figure}[tb]
    \centering
    \includegraphics[width=\linewidth]{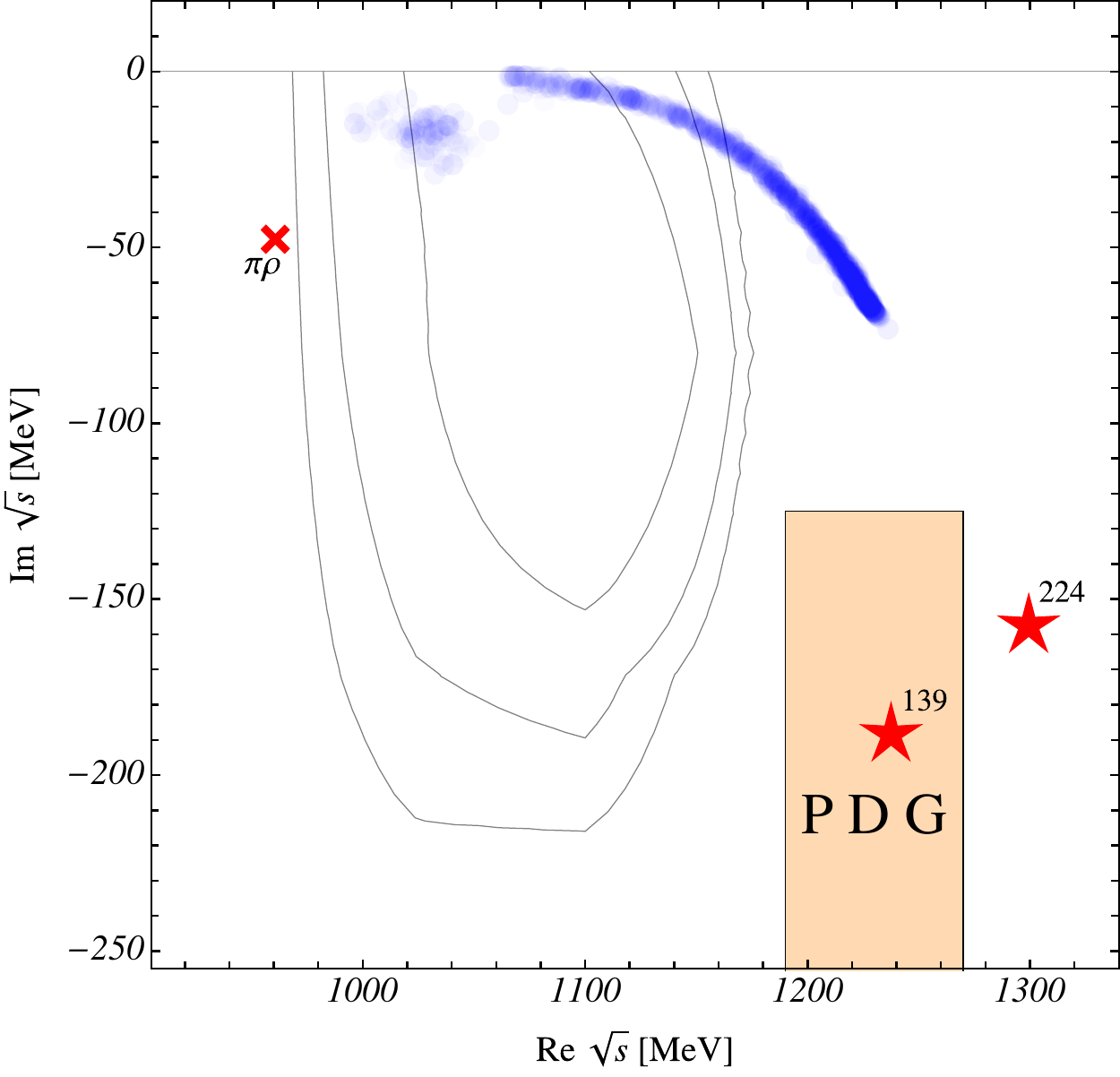}
    \caption{The $a_1$ pole positions from FVU (darker blue indicates higher sample density). The PDG result~\cite{ParticleDataGroup:2020ssz} and its uncertainties are included as the orange rectangle. The $\pi\rho$ branch point is indicated by the red cross and a naive chiral extrapolation with red stars (from $m_\pi=139$~MeV to $m_\pi=224$~MeV). The crude two-body Breit-Wigner/L\"uscher approximation is indicated with black contours.}
    \label{fig:A1-compilation}
\end{figure}

For each of the obtained parameter sets, we search for singularities of $T^c_{\ell'\ell}$ on the second Riemann sheet. The resulting pole positions are depicted in blue in Fig.~\ref{fig:A1-compilation}. As expected from the previous discussion of parameter correlation a precise determination of the $a_1$ pole position requires more input. Surprisingly, the distribution of poles is indeed finite with a stronger concentration around heavier $a_1$. This is apparent as the darker blue regions indicate higher sample density.

Putting our results into perspective:
1) We compare them to an approximate procedure employed earlier~\cite{Lang:2014tia}, assuming a stable $\rho$-meson. In that, using L\"uscher's method~\cite{Luscher:1986pf,Luscher:1990ux} the finite-volume spectrum is mapped to phase-shifts. Subsequently, a simple Breit-Wigner parametrization is used to determine the pole positions. The resulting confidence regions are depicted by the black (un-shaded) contours in Fig.~\ref{fig:A1-compilation}. It appears that this Breit-Wigner approach has only small overlap with the full FVU at lower masses, demonstrating the need for using the full three-body quantization condition;
2) We depict the current PDG values~\cite{ParticleDataGroup:2020ssz} as ${\sqrt{s}\approx M-i\,\Gamma/2}$ in Fig.~\ref{fig:A1-compilation}. The real part of the PDG mass overlaps with our predictions, but the PDG width is at least twice as large. This is expected since the pion mass in our case is heavier than the physical one, resulting in a reduced phase space for resonance decay;
3) We perform a chiral extrapolation of fits to experimental data~\cite{Sadasivan:2020syi}. The corresponding pole determination at the physical point is the first of its kind with a three-body unitary amplitude. Then, increasing the pion mass appearing in the loops and parameters of $K_n^{-1}$ only (see supplement~\ref{APP:projections}~\cite{supp} for technical details), we obtain the second red star in Fig.~\ref{fig:A1-compilation}. It confirms the expectation of the $a_1$ becoming heavier and narrower, although this does not lead to an overlap with the pole region from LQCD.

Finally, one can ask whether an explicit singularity in our parametrization leads to a bias towards the existence of an $a_1(1260)$. Removing that pole and allowing for one more term in the Laurent expansion, i.e., setting $C_{\ell'\ell}:=(c+c's)\delta_{\ell'0}\delta_{\ell 0}$,  one obtains fits that all lead to a pole in the $\pi\rho$ amplitude. While those poles are  concentrated close to the real axis at $\sqrt{s}\approx 1.04$~GeV, i.e., too light and too narrow, the exercise shows that $a_1$ poles are dynamically generated as demanded by LQCD data even if no explicit singularities are present in the parametrization of $C$.

\begin{figure}[b]
    \centering
    \includegraphics[width=\linewidth,trim=0 0.1cm 0 0,clip]{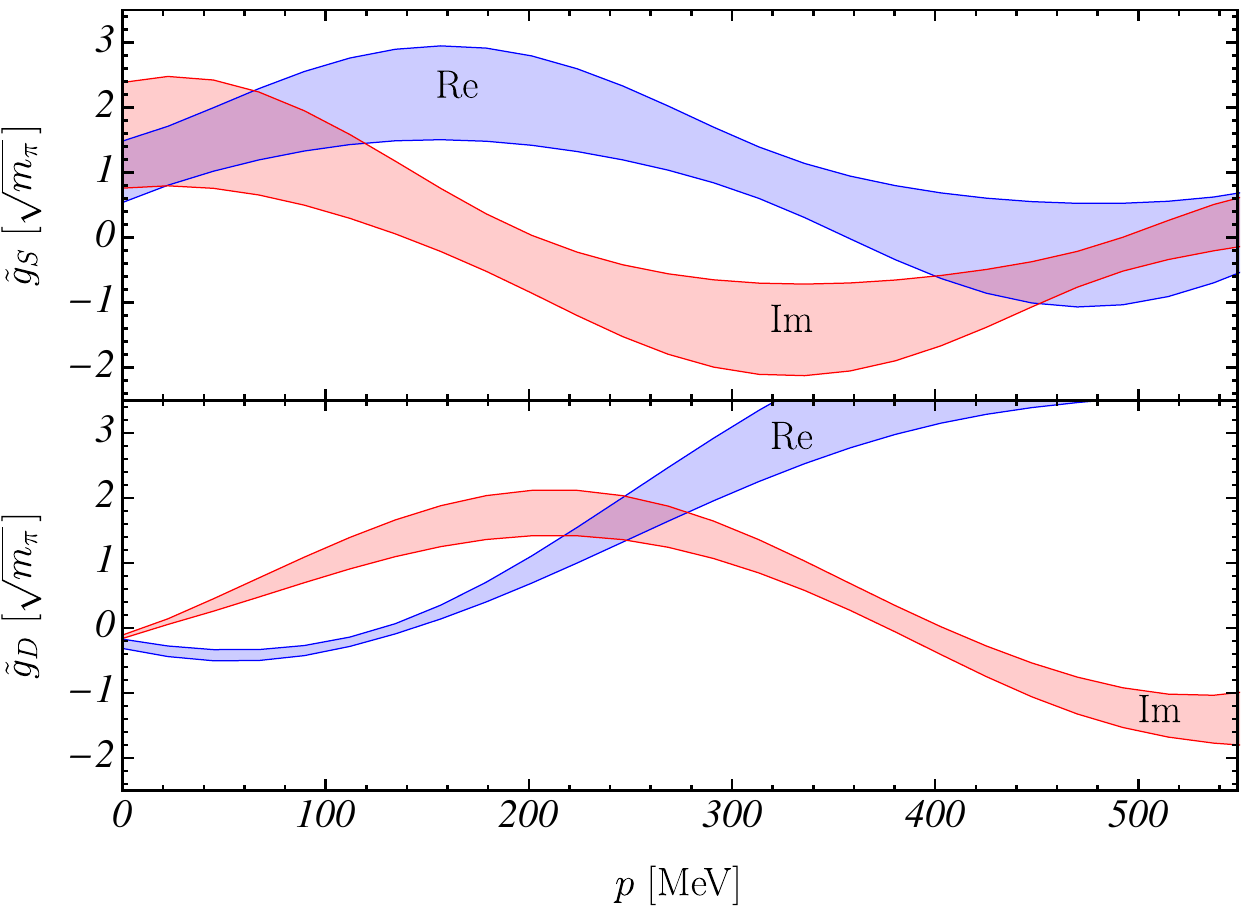}
    \caption{The 1-$\sigma$ confidence regions for the couplings $\tilde g_S$ and $\tilde g_D$ defined by pole residues (see text).}
    \label{fig:A1-couplings}
\end{figure}

The pole residues of the amplitude factorize~\cite{Sadasivan:2020syi}, ${\rm Res} (T^c_{\ell'\ell}(\sqrt{s}))=\tilde g_{\ell'}\tilde g_{\ell}$ in terms of couplings $\tilde g_S$ and $\tilde g_D$, analogously to the usual branching ratios but independent of background terms~\cite{ParticleDataGroup:2020ssz}. Their 1-$\sigma$ regions are shown in Fig.~\ref{fig:A1-couplings} as a function of real spectator momentum.
Clearly there are systematics attached (e.g. the missing $\pi\sigma$ channel) to this first determination of the resonance coupling, which can be addressed once the LQCD dataset is increased. The calculation of pole position and residues for a three-body unitary amplitude is another novelty of this work.

\bigskip
\noindent
\textit{Summary} ---
In this letter we have presented the first determination of the resonance parameters of the axial $a_1$-resonance from QCD. For that, three milestones had to be reached.
Firstly, the finite-volume spectrum for a resonant three-hadron system was determined including three-meson operators in a lattice QCD calculation.
Secondly, a three-body quantization condition including subsystems with spin and coupled channels was derived and applied to the finite-volume spectrum.
Finally, the corresponding unitary three-body scattering amplitude was solved and analytically continued to the complex plane to determine the pole positions and branching ratios.
We explored various forms of the short-range three-body force.
In our main solution we found an overlap of the mass of the $a_1$ with the phenomenological range, but substantially lower width.

This study paves the way for understanding exotic and hybrid resonances for which three-body dynamics are critical.
For the $a_1(1260)$ resonance, further extending the lattice calculation will have many benefits.
Additional data at this pion mass will resolve the sub-dominant channels like $\pi\sigma$, and lead to a more precise pole position of the $a_1(1260)$.
Results at other pion masses will help complete the picture of the $a_1$, its chiral trajectory, and its properties from first principles.


\bigskip
\noindent
\textit{Acknowledgments} ---
MM thanks Peter Bruns for useful discussions.
MD is grateful for the hospitality of the University of Valencia, where part of this work was done.
This material is based upon work supported by the National Science Foundation under Grant No. PHY-2012289 (MD,MM), the U.S. Department of Energy under Award Number DE-SC0016582 (MD,MM), DE-AC05-06OR23177 (MD,MM), and DE-FG02-95ER40907 (AA,FXL,RB,CC).
RB is also supported in part by the U.S. Department of Energy and ASCR, via a Jefferson Lab subcontract No. JSA-20-C0031. CC is supported by UK Research and Innovation grant {MR/S015418/1}.  Computations for this work were carried out in part on facilities of the USQCD Collaboration, which are funded by the Office of Science of the U.S. Department of Energy.
\bigskip
\bibliography{BIB}

\clearpage
\begin{onecolumngrid}
\setcounter{equation}{0}
\setcounter{figure}{0}
\setcounter{table}{0}
\setcounter{page}{1}
\makeatletter
\renewcommand{\theequation}{S\arabic{equation}}
\renewcommand{\thefigure}{S\arabic{figure}}
\setcounter{secnumdepth}{2}

\clearpage
\section*{\large Supplemental Material} \label{sec:sup}

\subsection{Lattice Operator Construction}
\label{APP:lat}
To extract multiple states in a finite-volume LQCD calculation, we use the variational method~\cite{Michael:1982gb,Luscher:1990ck,Blossier:2009kd} which requires building a large basis of interpolating operators with the same quantum numbers. To construct the operator basis for the $a_1(1260)$ we consider a simple $\bar{q}q$ interpolator and three types of multi-meson operators. These multi-meson operators are formed out of the $a_1(1260)$'s dominant decay channels, namely $\rho\pi$ and $\sigma\pi$, and the final decay state $\pi\pi\pi$.  To ensure our operators have the correct symmetry properties we begin by enforcing isospin matching the $a_1(1260)$, $I=1$, selecting the correct combinations of the charged $\rho$ and $\pi$.  The $I_3$ component is arbitrary in an $N_f=2$ simulation, and we choose $I_3=-1$.  The relevant combinations are given in Table~\ref{table:isospin_combinations}. Note that there are three different isospin configurations for the three-pion operator.  This stems from constructing a two-pion state with definite isospin~(with possiblities $0,1,2$), which we denote as $I^{(2)}$, and taking the tensor product of this state with a third pion.

\begin{table}[hbt]
    \centering
    \caption{Operators constructed to have definite isospin matching the $a_1(1260)$.  For the first two operators each meson $m$ is a function of it's momentum $\mathbf{p}_m$.  For the three-pion operators we distinguish the particles by indices $i=1,2,3$ where $\pi_i=\pi(\mathbf{p}_i)$.}
    \begin{ruledtabular}
    \begin{tabular}{c}
    Operators of Definite Isospin\\
    \hline
    \rule{0pt}{2.5ex}
    $\rho^-\pi^0 - \pi^-\rho^0 $ \\
    $\pi^-\sigma$ \\
    $6\pi^-_1\pi^-_2\pi^+_3+\pi^-_3\pi^-_2\pi^+_1+\pi^-_3\pi^-_1\pi^+_2+2\pi^-_3\pi^0_1\pi^0_2 -3\pi^-_2\pi^0_1\pi^0_3
    +3\pi^-_1\pi^0_2\pi^0_3 $\\
    $\pi^-_3\pi^-_2\pi^+_1-\pi^-_3\pi^-_1\pi^+_2-\pi^-_2\pi^0_1\pi^0_3+\pi^-_1\pi^0_2\pi^0_3$ \\
    $\pi^-_3\pi^-_2\pi^+_1+\pi^-_3\pi^-_1\pi^+_2-\pi^-_3\pi^0_1\pi^0_2$ \\
    \end{tabular}
    \end{ruledtabular}
    \label{table:isospin_combinations}
\end{table}

The last step in constructing the operator basis is to ensure these operators transform irreducibly under the $\mu$th row of irrep $\Lambda=T_{1g}$ of $O_h$. To do this we project each of our multi-meson operators as follows
\begin{equation}
\begin{aligned}
    \mathcal{O}^{\Lambda,\mu}_{\rho_i\pi}&=\frac{n_{\Lambda}}{\abs{G}}\sum_{g\in G}U^{\Lambda}_{\mu\mu}(g)\text{det}\left[R_g\right](R_g)_{ij}\rho_j(R_g \mathbf{p}_{\rho})\pi(R_g \mathbf{p}_{\pi})\,,\\
    \mathcal{O}^{\Lambda,\mu}_{\sigma\pi}&=\frac{n_{\Lambda}}{\abs{G}}\sum_{g\in G}U^{\Lambda}_{\mu\mu}(g)\text{det}\left[R_g\right]\sigma(R_g \mathbf{p}_s)\pi(R_g \mathbf{p}_{\pi})\,,\\
    \mathcal{O}^{\Lambda,\mu}_{(\pi\pi\pi)_{I^{(2)}}}&=\frac{n_{\Lambda}}{\abs{G}}\sum_{g\in G}U^{\Lambda}_{\mu\mu}(g)\text{det}\left[R_g\right]\left(\pi(R_g \mathbf{p}_1)\pi(R_g \mathbf{p}_2)\pi(R_g \mathbf{p}_3)\right)_{I^{(2)}}\,,
\end{aligned}
    \label{eqn:projectors}
\end{equation}
where $\mu$ is the row of the irrep, $n_{\Lambda}$ is the dimension of $\Lambda$, $U^{\Lambda}(g)$ is the $\Lambda$ representation matrix for group element $g$, and $R_g$ is the rotation matrix for group element $g$.

The extraction of the first excited state relied heavily on the inclusion of one particular operator, a three-pion operator with two of the pions having back to back momenta $\frac{2\pi}{L}(1,1,0)$, which we denote $\mathcal{O}^*_{(\pi\pi\pi)}$.  For this combination of momenta there is only one linearly independent isospin combination.  The effect on the excited state $E_2$ when including/excluding this operator can be seen in Fig.~\ref{fig:ostar_relevance}.  The fit result is depicted for different initial fit times $t_i$, while including/excluding this operator from the basis of $11$ operators.  Adding this operator significantly lowers the resulting energy of the state showing that the other operators in the basis do not have sufficient overlap with the state. Beyond $t_i=9$ we see that the energy level is stable and thus choose this fit time to extract the energy.

\begin{figure}[tbh]
    \centering
    \includegraphics[width=0.5\linewidth]{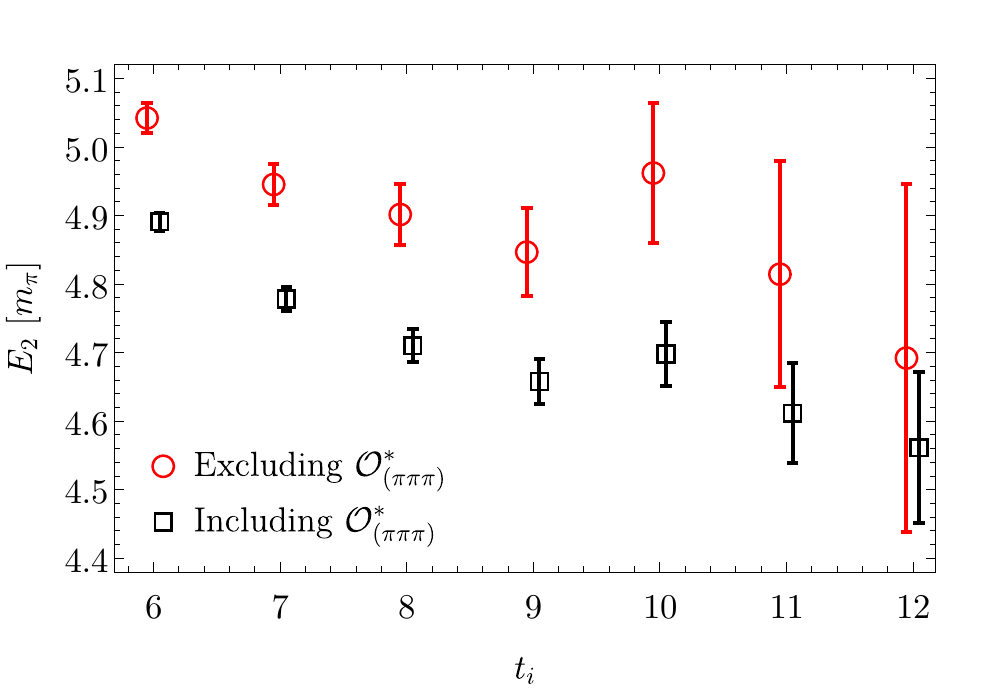}
    \caption{Plot of $E_2$ for different initial fit times $t_i$, with and without $O^*_{(\pi\pi\pi)}$.}
    \label{fig:ostar_relevance}
\end{figure}

\subsection{Technical details on spin-1 systems}
\label{APP:QC-details}

\subsubsection{Helicity formalism}

Here, we provide technical details on the implementation of spin-1 subsystems into the three-body setup of the FVU formalism~\cite{Mai:2017bge,Mai:2018djl}. In that, the generic form of the coupling of a spin-1 field to two asymptotically stable fields ($\rho\to\pi(p)\pi(q)$) is extracted from Ref.~\cite{Wess:1967jq}. Its spin-part reads
\begin{align}
\hat v(p,q)_\lambda=-i\varepsilon_\lambda^\mu(p+q)\,(p_\mu-q_\mu) \,.
\end{align}
Note that the isospin part is taken care of in the main definitions of the letter, where also the coupling constant was reabsorbed. For each helicity state $\lambda\in\{-1,0,+1\}$ of the spin-1 field, the four-vector $\varepsilon$ depends on the direction of the propagation~\cite{Chung:1971ri} as
\begin{align}
\varepsilon_0(\bm{p})=\frac{1}{m_\rho}\begin{pmatrix}p\\E^\rho_{p}\hat{\bm{p}}_x\\E^\rho_{p}\hat{\bm{p}}_y\\E^\rho_{p}\hat{\bm{p}}_z\end{pmatrix}\,,
~
\varepsilon_{\pm 1}(\bm{p})=\frac{1}{\sqrt{2}}\begin{pmatrix}
0 \\
\mp \cos{\theta_{\hat{\bm{p}}}}\cos{\phi_{\hat{\bm{p}}}}+i\sin{\phi_{\hat{\bm{p}}}} \\
\mp \cos{\theta_{\hat{\bm{p}}}}\sin{\phi_{\hat{\bm{p}}}}-i\cos{\phi_{\hat{\bm{p}}}} \\
\pm \sin{\theta_{\hat{\bm{p}}}} \end{pmatrix}\,,
\label{eq:helicity-vectors}
\end{align}
where $E^\rho_p:=\sqrt{m_\rho^2+\bm{p}^2}$. On-shell, this fulfills required properties, such as the transversality, i.e. $p_\mu\varepsilon^\mu=0$ exactly, see Ref. \cite{Chung:1971ri}. Away from the on-shell point one can generalize the above definitions using $m_\rho\to\sqrt{E_p^2-p^2}$. However, as the difference between both versions does not lead to new singularities of the spin-1 propagator, perturbation theory is viable, allowing one to reabsorb it into the local terms~\cite{Bruns:2013tja}.

As discussed in the main text, evaluating the self-energy term ($\Sigma$) in a finite volume ($\Sigma^L$) boils down to replacing the three-dimensional integration to a summation over meson momenta in the three-body center of mass $\mathcal{S}_L:=(2\pi/L)\mathds{Z}^3$
\begin{align}
\Sigma^L_{n,\lambda'\lambda}&(s,\bm{p})=
\frac{J(\bm{p})}{L^3}
\sum_{\bm{k}\in \mathcal{S}_L}
\frac{\sigma_p^n}{(4E_{k^\star}^2)^n}
\frac{
\epsilon_{\lambda'}^{\star\nu*}(\bm{P}_3-\bm{p})(P_\nu^\star-p_\nu^\star-2k_\nu^\star)
\epsilon_{\lambda}^{\star\mu}(\bm{P}_3-\bm{p})(P_\mu^\star-p_\mu^\star-2k_\mu^\star)
}{ 2E_{k^\star}(\sigma_p-4E_{k^\star}^2)}\,.
\end{align}
All elements are evaluated in the two-body reference frame, marked by $(\star)$. Expressing those in term of the three-body center of mass momenta calls for the use of the following boost
\begin{align}
\label{eq:boost}
\bm{k}^\star(s,\bm{k},\bm{p})
=
\bm{k}+
\bm{p}\left(\frac{\bm{k}\cdot\bm{p}}{\bm{p}^2}\left(J(s,\bm{p})-1\right)+
\frac{1}{2}J(s,\bm{p})\right)\,,
\quad
J(s,\bm{p})=\frac{\sqrt{\sigma_{p}}}{\sqrt{s}-E_{\bm{p}}}\,,
\end{align}
while the helicity state vectors are given by
\begin{align}
\varepsilon^\star_0(\bm{p})=\begin{pmatrix}0\\\hat{\bm{p}}_x\\\hat{\bm{p}}_y\\\hat{\bm{p}}_z\end{pmatrix}\,,
~
\varepsilon^\star_{\pm 1}(\bm{p})=\frac{1}{\sqrt{2}}\begin{pmatrix}
0 \\
\mp \cos{\theta_{\hat{\bm{p}}}}\cos{\phi_{\hat{\bm{p}}}}+i\sin{\phi_{\hat{\bm{p}}}} \\
\mp \cos{\theta_{\hat{\bm{p}}}}\sin{\phi_{\hat{\bm{p}}}}-i\cos{\phi_{\hat{\bm{p}}}} \\
\pm \sin{\theta_{\hat{\bm{p}}}} \end{pmatrix}\,.
\end{align}
Finally we note that boosts, and with it the above prescription, are not well defined for off-shell states. This technical challenge can be resolved by simply noting that in the unphysical region, finite- and infinite-volume self-energy terms coincide up to $e^{-m_\pi L}$ effects. Thus, we replace
\begin{align}
\Sigma^L(s)\mapsto\Sigma(s) &\text{~~for~~} \sqrt{s}<W_{\rm reg+} + 0.1(W_{\rm phys} - W_{\rm reg+})\,,
\end{align}
where $W_{\rm reg\pm}=E_p\pm p$ and $W_{\rm phys}=\sqrt{\bm{p}^2+4m_\pi^2}+E_p$. Numerical checks were performed to ensure negligible dependence on the choice of matching point, see the discussion in Refs.~\cite{Mai:2018djl,Brett:2021wyd}.

\subsubsection{Partial-wave Projection}

The central object of the infinite-volume calculation is the partial-wave projected coupled-channel amplitude in the $JLS$ basis,  $T^c_{\ell'\ell} (s,p',p)$, where ${\ell^{(\prime)}\in\{S,D\}}$ and $p^{(\prime)}=|\bm{p}^{(\prime)}|$. It is obtained from the plane-wave helicity amplitude $T^c_{\lambda'\lambda} (s,\bm{p}',\bm{p})$ of the main text Eq.~\eqref{eq:T3-integral-equation}, where $\lambda^{(\prime)}\in\{-1,0,1\}$, by a standard partial-wave decomposition~\cite{Sadasivan:2020syi}. In summary,
\begin{align}
{\cal A}_{\lambda'\lambda} (s,\bm{p}',\bm{p})&=
\sum_{M=-J}^J\frac{2J+1}{4\pi}\,
\mathfrak{D}_{M\lambda'}^{J*}(\phi_{\bm{p}^\prime},\theta_{\bm{p}^\prime},0)\,
{\cal A}_{\lambda'\lambda}^J(s,{p}^\prime,{p})\,\mathfrak{D}_{M\lambda}^{J}(\phi_{\bm{p}},\theta_{\bm{p}},0)\,,\\
{\cal A}_{\lambda'\lambda}^J(s,{p}^\prime,{p})&=
U_{\lambda'\ell'}{\cal A}_{\ell'\ell}(s,p',p)U_{\ell\lambda}\,,\label{eq:tlpl}\\
U_{\ell\lambda}&:=\sqrt{\frac{2\ell+1}{2J+1}}(\ell 01\lambda|J\lambda)(1\lambda00|1\lambda))=
\left(
\begin{array}{ccc}
 \frac{1}{\sqrt{3}} & \frac{1}{\sqrt{3}} & \frac{1}{\sqrt{3}}\,,\\
 \frac{1}{\sqrt{6}} & -\sqrt{\frac{2}{3}} & \frac{1}{\sqrt{6}}\,,\\
\end{array}
\right)\,,
\end{align}
where $J=1$ is the $a_1$ spin, the $(\dots|\dots)$ indicate Clebsch-Gordan coefficients, $\mathfrak{D}$ are Wigner-D functions defined to be consistent with the rotations of the polarization vectors of Eq.~\eqref{eq:helicity-vectors}, in the same convention as in Ref.~\cite{Chung:1971ri}, and ${\cal A}\in \{T^c, B, C\}$ stands for the connected $T$-matrix, pion exchange term, and contact term, respectively, as defined in the main text.

\subsection{Fit details}
\label{APP:statistics}

In this section, we provide some additional details about the distribution of parameters in the central, four-parameter fit discussed in the main part of the manuscript.
The free parameters of the fit are $\{m_{a_1},g_S,g_D,c\}$ which are determined by minimizing the correlated $\chi^2$ with respect to the two energy eigenvalues and covariance matrix thereof ($\tilde C$),
\begin{align}
\begin{pmatrix}E_1\\E_2\end{pmatrix}=
\begin{pmatrix}4.2956\\4.6575\end{pmatrix}\,m_\pi
\text{~~and~~}
\tilde C=10^{-3}
\begin{pmatrix}
0.984717& 0.20161\\
0.201609& 1.04927
\end{pmatrix}\,m_\pi^2\,.
\end{align}

In our first investigations we found that $E_1$ and $E_2$ are most sensitive to $\{m_{a_1},g_D\}$ and $\{g_S,c\}$, respectively. Quantitatively, this is confirmed by re-sampling $\{E_1,E_2\}$ and fitting that sample with a randomized starting value for $m_{a_1}$. The resulting distribution of $\chi^2$-values and parameters, as well as pairwise correlations between them are depicted in Fig.~\ref{fig:Chi2-histogram}. We distinguish a small set (72/2000) of ``exotic'' solutions in Fig.~\ref{fig:A1-compilation} of the main text (``Cluster 1'') which lead to pole positions clustered at around $\Re \sqrt{s}\approx 1040$~MeV in order to study their origin.

First, we note that albeit a clear overfit, the $\chi^2$-distribution follows a very natural behavior. Also the solutions of ``Cluster 1'' do not exhibit any unusual behavior like order-of-magnitude different fit parameters. Similarly, neither in terms of the bare mass $m_{a_1}$ nor $g_D$ it appears to be distinguishable from the remainder of the fits. However, it appears that the vast majority of exotic solutions has both large $c$ and $g_S$. This can be understood insofar as both these values can compensate each other to a large extend with respect to the position of the ground level. However, too large parameter values lead to a rather small $a_1$ mass. In summary solutions in ``Cluster 1'' cannot be excluded at the moment. Given the large correlations between the corresponding bare parameters, it can be expected that a larger data base can indeed help ruling out such solutions. At the moment we can only denote the observation that Cluster 1 disappears if the four-parameter fit is reduced to a three-parameter fit with $c=0$, i.e., a pure resonance fit without any additional constant term.

\begin{figure}[t]
    \centering
    \includegraphics[width=\linewidth]{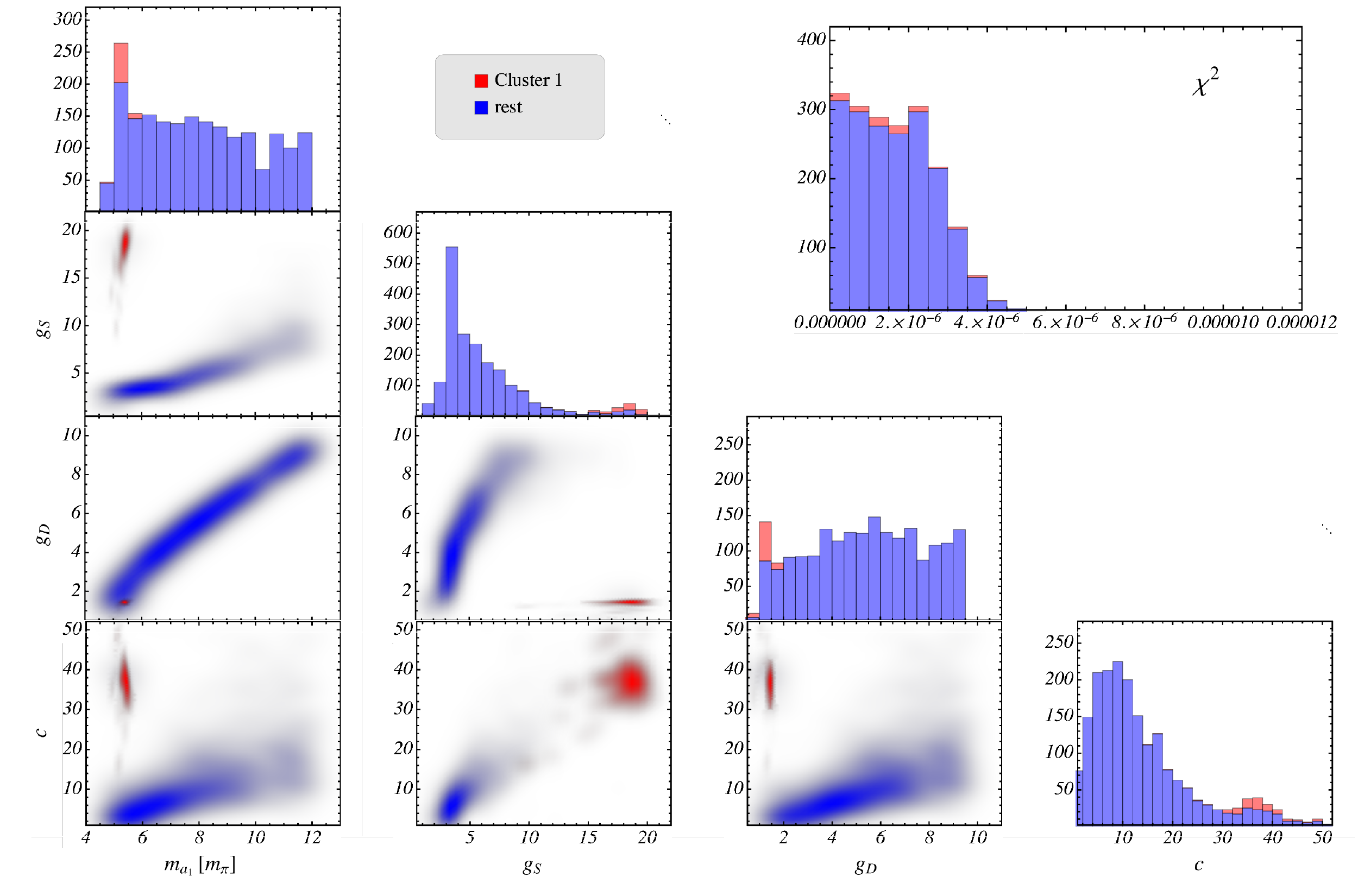}
    \caption{Details of the four-parameter fit to the LQCD energy eigenvalues. Top right inset shows the distribution of the total $\chi^2$, while the rest shows parameter distribution and smoothed pairwise correlations. ``Cluster 1'' refers to a pole region at around $\sqrt{s}\approx 1040$~MeV close to the real $\sqrt{s}$ axis that is almost disjoint from the ``rest'', i.e., the curved region in Fig.~\ref{fig:A1-compilation} of the main text.
    }
    \label{fig:Chi2-histogram}
\end{figure}

\subsection{Analytic continuation, and chiral extrapolation}
\label{APP:projections}

\subsubsection{Analytic continuation}

The coupled-channel partial-wave amplitude can be expanded in $\sqrt{s}$ around the pole position $\sqrt{s_0}$ of the $a_1(1260)$,
\begin{align}
T^c_{\ell'\ell} (s,p',p)=\frac{\tilde g_{\ell'}\, \tilde g_\ell}{\sqrt{s}-\sqrt{s_0}}+{\cal O}(1) \,,
\end{align}
with $\tilde g_\ell$ playing the role of (Breit-Wigner) branching ratios, but defined at the pole~\cite{ParticleDataGroup:2020ssz}. In addition, for the current case of $\pi\rho$ scattering in $S$ and $D$-waves, $\tilde g$ are necessarily functions of spectator momentum, $\tilde g_\ell\equiv\tilde g_\ell(p)$ and $\tilde g_{\ell'}\equiv\tilde g_{\ell'}(p')$. Analogously one might think of resonance transition form factors that are pole residues depending on photon virtuality~\cite{Kamano:2018sfb, Mai:2021vsw}. For a numerically stable method to calculate residues see Appendix C of Ref.~\cite{Doring:2010ap}. We show the $\tilde g_\ell(p)$ for real spectator momenta $p$ in Fig.~\ref{fig:A1-couplings}, which requires another analytic extrapolation from the complex $p$ on the spectator momentum contour (SMC) at which the solution is calculated (see next subsection).


As for the analytic continuation to find the $a_1(1260)$ in the first place, one has to realize that the analytic structure of the three-body amplitude is considerably more complicated than for two-body amplitudes~\cite{Doring:2009yv}. In Ref.~\cite{Doring:2009yv} an approximate method for the continuation was proposed and applied, but here we sketch the exact solution newly developed in this work. Summarizing some results from Ref.~\cite{Doring:2009yv}, the $3\pi$ amplitude with a resonant $\rho$ subsystem develops three branch points: one at the real threshold $s=(3m_\pi)^2$ and two at the complex $\pi\rho$ thresholds at $\sqrt{s}=\sqrt{\sigma_\rho}+m_\pi$
and $\sqrt{s}=(\sqrt{\sigma_\rho})^*+m_\pi$, where $\sqrt{\sigma_\rho}$ is the $\rho$ pole position expressed in the $\pi\pi$ invariant mass $\sqrt{\sigma}$. The complex branch points lie on the second Riemann sheet that is accessed from the physical axis by analytically continuing through the cut of the $s=(3m_\pi)^2$ branch point into the lower $\sqrt{s}$ half-plane. The complex branch points induce one additional Riemann sheet, each (sheets 3 and 4), and it is a matter of choice on which sheet 2, 3, or 4 to search for poles. Here, the situation is clear and the $a_1(1260)$ is on sheet 2 because its pole lies well above the complex branch points in $\sqrt{s}$ (see Fig.~\ref{fig:A1-compilation} of the main text). Access to the different sheets is enabled by suitable contour deformation, both for the spectator momentum integration (SMC) and self-energy integration (SEC), see Eqs.~\eqref{eq:T3-integral-equation} and \eqref{eq:tau-infinite}, respectively.
\begin{figure}
    \centering
    \includegraphics[width=0.48\linewidth]{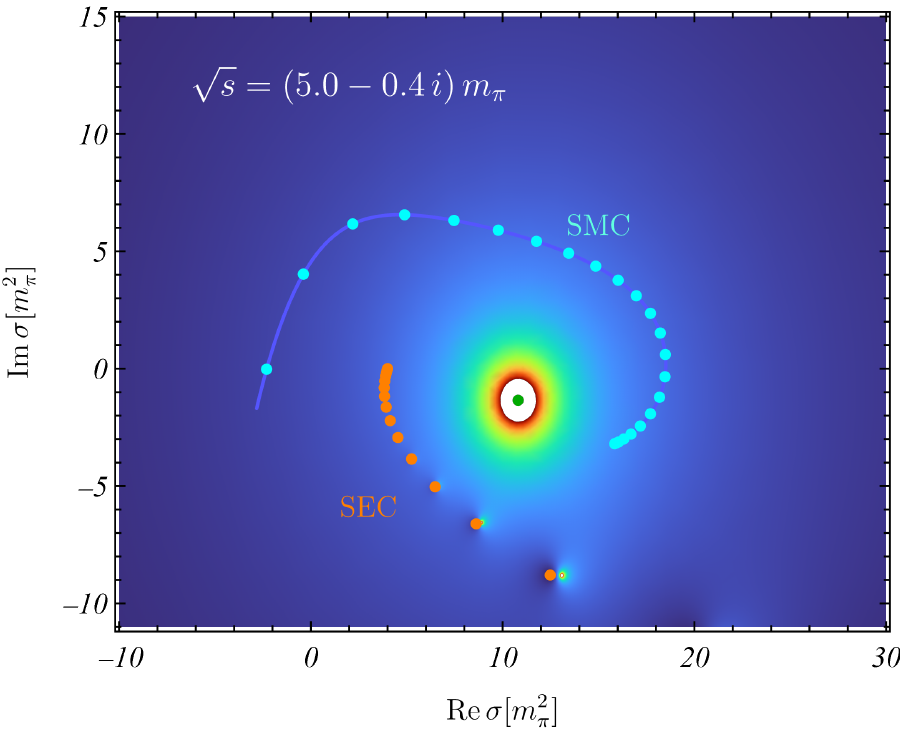}
    ~~
    \includegraphics[width=0.47\linewidth]{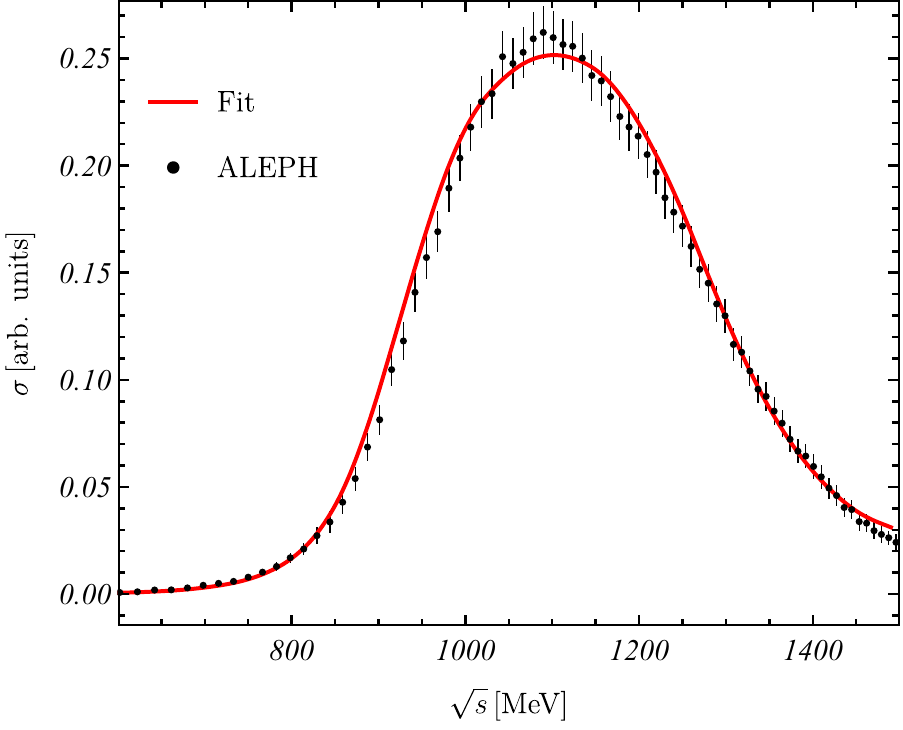}
    \caption{{\bf Left}: Example for SMC and SEC contours in the complex plane of the two-body energy-squared, $\sigma$. Typical Gauss point distributions on the contours are indicated with turquoise and orange dots, respectively. The pole of the $\rho$ is highlighted, as well (green circle). The topology of contours and $\rho$ pole shown in this figure corresponds to being on the second Riemann sheet of $T^c_{\ell'\ell}$ where the $a_1(1260)$ pole is found. {\bf Right}: Fit of the three-body unitary $\pi\rho$ amplitude (red solid line) to the line shape data measured by the ALEPH collaboration~\cite{Schael:2005am}.
    }
    \label{fig:integrations}
\end{figure}
In Fig.~\ref{fig:integrations} to the left a typical case is shown, for a complex three-body energy of $\sqrt{s}=(5.0-0.4i)\,m_\pi$. In short, the contours are constructed to 1) avoid each other, 2) avoid the $\rho$ pole, 3) avoid three-body cuts of the $B$-term, and 4) pass the $\rho$ pole to the left or to right, to access different Riemann sheets. By consequently utilizing these construction principles, one can access all Riemann sheets exactly. For more details, see Ref.~\cite{Doring:2009yv}.

\subsubsection{Chiral extrapolation}
There is no model for the chiral extrapolation of the $a_1(1260)$ as rigorous as in the case of (coupled-channel) meson-meson scattering, where chiral amplitudes to a given order can be matched to unitarity by the Inverse Amplitude or similar methods~\cite{Truong:1988zp, Dobado:1996ps, Mai:2019pqr, Culver:2019qtx, Guo:2018zss, Hu:2017wli, Doring:2016bdr, Hu:2016shf, Bolton:2015psa}. However, effective Lagrangians exist~\cite{Birse:1996hd} that have been used to generate the $a_1(1260)$ dynamically~\cite{Lutz:2003fm, Roca:2005nm}. Here, we simply adopt the three-body unitary, infinite-volume amplitude $T^c_{\ell'\ell}$ from Eq.~\eqref{eq:tlpl} and use the methods of Ref.~\cite{Sadasivan:2020syi} to fit the free parameters to physical $\pi\pi$ phase shifts and to the $\pi\pi\pi$ lineshape measured by the ALEPH collaboration~\cite{Schael:2005am}, with the result shown in Fig.~\ref{fig:integrations} to the right. Notably, the fit is much better than in Ref.~\cite{Sadasivan:2020syi} due to an automated fit procedure and better capturing of the two-body input. Subsequently the pole position is extracted, which represents the very first pole determination from experiment with a three-body unitary amplitude. The result lies within the intervals quoted in the PDG (see Ref.~\cite{ParticleDataGroup:2020ssz} and Fig.~\ref{fig:A1-compilation} of the main text).

We can then qualitatively study the chiral trajectory of the amplitude and $a_1(1260)$ pole by changing the pion mass in $\tau$ and $B$ from Eqs.~\eqref{eq:tau-infinite} and \eqref{eq:Bterm}, respectively. In addition, we use the $\pi\pi$ amplitude in the $\rho$-channel for the unphysical pion mass from Ref.~\cite{Mai:2019pqr}, as done for all other parts of this letter. However, apart from these changes we can only assume that all other parameters, including the spectator momentum cutoff $\Lambda$, do not depend on the quark mass because no theory exists for these parameters. To reiterate, this is an insufficient approximation, but it provides us an order-of magnitude estimate of how the $a_1(1260)$ could change. Indeed as Fig.~\ref{fig:A1-compilation} of the main text shows, the resonance gets narrower and heavier, as expected from general phase space considerations.

\end{onecolumngrid}
\end{document}

%% file: chris_definitions.tex
\usepackage{mleftright,xparse}
\NewDocumentCommand\xDeclarePairedDelimiter{mmm}
{%
	\NewDocumentCommand#1{som}{%
		\IfNoValueTF{##2}
		{\IfBooleanTF{##1}{#2##3#3}{\mleft#2##3\mright#3}}
		{\mathopen{\csname##2\endcsname#2}##3\mathclose{\csname##2\endcsname#3}}%
	}%
}
\xDeclarePairedDelimiter{\av}{\langle}{\rangle}
\xDeclarePairedDelimiter{\abs}{\lvert}{\rvert}
\xDeclarePairedDelimiter{\ket}{|}{\rangle}
\xDeclarePairedDelimiter{\bra}{\langle}{|}
\NewDocumentCommand\braket{somm}{%
	\IfNoValueTF{#2}{\mleft\langle #3\,|#4\mright\rangle}{NOTIMPLEMENTED}
}
\NewDocumentCommand\opbraket{sommm}{%
	\IfNoValueTF{#2}
	{\IfBooleanTF{#1}{\langle#3|#4|#5\rangle}{\mleft\langle #3 \left| #4 \right| #5 \mright\rangle}}
	{\mathopen{\csname#2\endcsname\langle}#3\mathopen{\csname#2\endcsname|} #4 \mathclose{\csname#2\endcsname|} #5\mathclose{\csname#2\endcsname\rangle}}
}

\def\beq{\begin{equation}}
\def\eeq{\end{equation}}
\def\beqs#1\eeqs{\beq\begin{split} #1 \end{split}\eeq}

\usepackage{tabularx}
\usepackage{booktabs}
\usepackage{graphicx}
\newcolumntype{L}{>{$}l<{$}}
\newcolumntype{C}{>{$}c<{$}}
\newcolumntype{R}{>{$}r<{$}}

\AtBeginDocument{
\heavyrulewidth=.08em
\lightrulewidth=.05em
\cmidrulewidth=.03em
\belowrulesep=.65ex
\belowbottomsep=0pt
\aboverulesep=.4ex
\abovetopsep=0pt
\cmidrulesep=\doublerulesep
\cmidrulekern=.5em
\defaultaddspace=.5em
}

\NewDocumentCommand{\nicetablespace}{ O{10pt} O{1.2} }
{
    \setlength{\tabcolsep}{#1}
    \renewcommand{\arraystretch}{#2}
}

\usepackage{dsfont}

\usepackage{enumitem}

\usepackage[normalem]{ulem}   